\documentclass[aps,preprint]{revtex4}
\usepackage{amsfonts}
\usepackage{amsmath}
\usepackage{amssymb}
\usepackage{epsf}
\usepackage{epsfig}
\usepackage{graphicx}
\usepackage{rotating}
\usepackage{boxedminipage}

\begin{document}

\title{Revisiting the critical behavior of nonequilibrium models in
short-time Monte Carlo simulations}
\author{Roberto da Silva}
\email{rdasilva@inf.ufrgs.br (corresponding author)}
\affiliation{Departamento de Inform\'{a}tica Te\'{o}rica, Instituto de Inform\'{a}tica,
Universidade Federal do Rio Grande do Sul. \\
Av. Bento Gon\c{c}alves, 9500, CEP 90570-051 Porto Alegre RS Brazil}
\author{Ronald Dickman}
\email{dickman@uari.if.ufmg.br}
\affiliation{Departamento de F\'{\i}sica, Instituto de Ci\^{e}ncias Exatas, Universidade
Federal do Rio Grande do Sul\\
Av. Ant\^{o}nio Carlos 6627, CEP 30123-970\\
Belo Horizonte, Minas Gerais, Brazil}
\author{J. R. Drugowich de Fel\'{\i}cio}
\email{drugo@usp.br}
\affiliation{Departamento de F\'{\i}sica e Matem\'{a}tica, Faculdade de Filosofia, Ci\^{e}%
ncias e Letras, Universidade de S\~{a}o Paulo.\\
Av. Bandeirantes, 3900-CEP 014040-901 Ribeir\~{a}o Preto SP Brazil}

\begin{abstract}
We analyze two alternative methods for determining the exponent $z$ of the
contact process (CP) and Domany-Kinzel (DK) cellular automaton in Monte
Carlo Simulations. One method employs mixed initial conditions, as proposed
for magnetic models [Phys. Lett. A. \textbf{298}, 325 (2002)]; the other is
based on the growth of the moment ratio $m(t)=\langle \rho ^{2}(t)\rangle
/\langle \rho (t)\rangle ^{2}$ starting with all sites occupied. The methods
provide reliable estimates for $z$ using the short time dynamics of the
process. Estimates of $\nu _{||}$ are obtained using a method suggested by
Grassberger.
\end{abstract}

\maketitle

\setlength{\baselineskip}{0.7cm}

\section{Introdution}


The dynamics of spin models quenched from high temperature to the critical
point is a subject of considerable current interest, because the initial
phase of the relaxation (the so-called short time dynamics) carries
important information about static and dynamic critical behavior. Using a
phenomenological renormalization group analysis, Janssen, Schaub and
Schmittmann \cite{Janssen} demonstrated the scaling law 
\begin{equation}
\left\langle M\right\rangle (t)\sim m_{0}t^{\theta }  \label{stM}
\end{equation}
characterized by the new critical exponent $\theta $. (Here $\left\langle
\cdot \right\rangle $ denotes an average over initial configurations
consistent with initial magnetization $m_0$, and over the noise in the
stochastic dynamics.) Relation (\ref{stM}) holds for times $t < t_{\max
}\sim m_{0}^{-z/x_{0}}$, where $z$ is the dynamic exponent, while $m_{0}$
denotes the initial magnetization, and $x_{0} $ its anomalous dimension. The
``critical initial slip" described by Eq.~(\ref{stM}) emerges from an
initially disordered state, and is characteristic of a stochastic, far from
equilibrium relaxation process. This approach also offers a way to determine
the \textit{static} critical exponents, since the scaling law used to deduce
Eq. (\ref{stM}) can also be applied to higher moments of the order
parameter, yielding: 
\begin{equation}
\left\langle M^{k}\right\rangle (t,m_{0})=L^{-\frac{k\beta }{\nu }%
}\left\langle M^{k}\right\rangle (L^{-z}t,L^{x_{0}}m_{0})
\end{equation}

Models with absorbing states exhibit scaling behavior at the critical point
marking the transition between active and absorbing stationary states, even
though the stationary state is not given by a Boltzmann distribution \cite%
{Hinrichsen,Dickman1}. The order parameter is the activity density $\rho
\geq 0$, given by $\rho (t)=L^{-d}\sum_{i=1}^{L^{d}}\sigma _{i}\geq 0$,
where $\sigma _{i}= 1$ (0) corresponds to the presence (absence) of activity
at site $i$ ($d$ is the dimensionality of the system). Activity is commonly
associated with the presence of a ``particle". For such models the following
scaling law \cite{Hinrichsen} has been conjectured: 
\begin{equation}
\rho (t)\sim t^{-\beta /\nu _{||}}f\left[ (p-p_{c})t^{1/\nu
_{||}},t^{d/z}/N,\rho _{0}t^{\beta /\nu _{\shortparallel }+\theta }\right] 
\text{,}  \label{density}
\end{equation}%
where $\rho _{0}$ is the initial density, $p$ is a temperaturelike control
parameter, $p_c$ is its critical value, and $N=L^{d}$ is the number of
sites. The exponent $\beta $ is associated with dependence of the stationary
value of $\rho$ on the control parameter: $\rho\sim (p-p_{c})^{\beta }$,
while $\nu _{||}$ and $\nu _{\perp }$ are the critical exponents associated
with the correlation time ($\xi _{||}\sim \left\vert p-p_{c}\right\vert
^{-\nu _{_{||}}}$) and correlation length ($\xi _{\perp }\sim \left\vert
p-p_{c}\right\vert ^{-\nu _{\perp }}$). Given the defining relation $\xi
_{||}\sim \xi _{\perp }^{z}$, the dynamic exponent $z=\nu_{_{||}}/\nu
_{\perp }$. Letting $p \to p_c$ and $N \to \infty$, we expect the scaling
function $f$ in Eq. (\ref{density}) to have the property: 
\begin{equation}
f\left[ 0,0,u\right] =\left\{ 
\begin{tabular}{ll}
$u$, & $u \simeq 0$ \\ 
\  & \  \\ 
$C$, & $u\rightarrow \infty $%
\end{tabular}%
\ \right.
\end{equation}%
where $C$ is a constant.

In a realization of the process at criticality ($p=p_c$) beginning with a
completely filled lattice $(\rho _{0}=1$), the activity density decays via
the power law $\rho (t)\sim t^{-\beta /\nu _{_{||}}}$ for $t \ll L^z$. On
the other hand, in realizations starting with only a single particle or
active site (spreading process), the average number of particles increases
with time: $\rho (t)\sim t^{\eta }$, where $\eta =(d\nu _{\perp }-2\beta
)/\nu _{_{||}}$.

An interesting crossover phenomenon in the evolution of $\rho (t),$ between
the initial increase $(\sim t^{\eta }$) asymptotic decay ($\sim t^{-\beta
/\nu _{_{||}}})$, emerges in a critical spreading process with low initial
particle density. The crossover time $t_{c}$ is related to the initial
density $\rho _{0}$ via

\begin{equation}
t_{c}\sim \rho _{0}^{-1/(\beta /\nu _{_{||}}+\eta )}\text{.}
\end{equation}
A process starting with a single particle ($\rho _{0}=1/L$) corresponds to $%
\rho _{0}\rightarrow 0$ for large lattices, so that $t_{c}$ diverges and the
spreading regime $\rho (t)\sim t^{\eta }$ extends over the entire evolution.

A method for determining the critical exponent $z,$ proposed in the context
of the short time behavior of magnetic models \cite{da Silva}, suggests that
we consider the function

\begin{equation}
F_{2}(t)=\frac{\;\;\left\langle \rho \right\rangle _{\rho _{0}=1/L^{d}}}{%
\left\langle \rho \right\rangle _{\rho _{0}=1}^{2}}  \label{second_cumun}
\end{equation}%
which has the asymptotic behavior $F_{2}(t)\sim t^{d\frac{\nu _{\perp }}{\nu
_{_{||}}}}=t^{d/z}$. Thus we can obtain $z$ directly by analyzing short time
simulation results with mixed initial conditions.

In this paper we employ this approach to calculate $z$ for the contact
process (CP) and the Domany-Kinzel cellular automaton (DK), both known to
belong to the universality class of directed percolation\cite{Dickman1}. We
should note that in the literature on absorbing-state phase transitions $z $
is commonly used to denote the exponent governing the growth of the
mean-square distance of particles from the original seed, in spreading
simulations. To avoid confusion we denote the latter exponent as $Z$, so
that $R^{2}\sim t^{Z}$. The dynamic exponent is then related to the
spreading exponent via $z=2/Z$. 

A method to estimate the exponent $\nu _{||}$ was suggested by Grassberger 
\cite{grassberger}, who used the relation 
\begin{equation}
D(t)\equiv \left. \frac{\partial \ln \rho }{\partial p }\right\vert
_{p=p_c}\sim t^{1/\nu z}  \label{powerlawderivative}
\end{equation}%
where in a simulation the derivative is evaluated numerically via 
\begin{equation}
D(t)=\frac{1}{2h}\ln \left( \frac{\rho (p_{c}+h)} {\rho (p_{c}-h)}\right) 
\text{,}  \label{numeric derivative}
\end{equation}
which evidently requires data for values of $p$ slightly off critical. In
this context a reweighting scheme that permits one to study various values
of $p$ in the same simulation is particularly convenient \cite{Dickman2}.

In the following section we present details on the models and our simulation
technique. In section III we report and analyze the simulation results, and
in section IV present our conclusions.

\section{Models and simulation method}

The contact process was introduced by Harris as a toy model of epidemic
propagation. It evolves in continuous time. Denoting the number of occupied
nearest neighbors of site $i$ by $n_{i}=\sum_{j\in \left\langle
i\right\rangle } \sigma _{j}$, where $\left\langle i\right\rangle $ denotes
the set of nearest neighbors of site $i$, the transition rates are

\begin{equation}
\begin{tabular}{lll}
$p(0\rightarrow 1,n)$ & $=$ & $\frac{\lambda n}{2d} \;\;\;\;\; $(creation)
\\ 
\  & \  & \  \\ 
$\ p(1\rightarrow 0,n)$ & $=$ & $1 \;\;\;\;\;$ (annihilation)$.$%
\end{tabular}
\label{regras contato}
\end{equation}%
The model suffers a continuous transition at a critical value $\lambda_c$;
in one dimension $\lambda _{c}=3.29785(8)$ \cite{Dickman1}.

As described in Refs. \cite{Dickman1,Dickman2}, our simulations employ a
list of occupied sites and sample reweighting to improve efficiency. Since
the choice of sites in the dynamics is restricted to the occupied set, the
time increment $\Delta t$ associated with each event (annihilation or
creation) is $1/N_{p}$, where $N_{p}$ is the number of occupied sites
immediately prior to the event. A given realization of the process ends at a
predetermined maximum time, or when all particles have been annihilated.
Reweighting is used to study the effects of window size $h$ in Eq. (\ref%
{numeric derivative}), in determining $\nu_{||}$ for the CP, using $\rho
_{0}=1/L$.

The Domany Kinzel (DK) cellular automaton is a discrete-time process
exhibiting a phase transition between an active and an absorbing phase of
the same kind as in the CP \cite{domany}. Each site of the lattice can be in
one of two states, $\sigma _{i}=1$(active) or $\sigma _{i}=0$ (inactive).
The transition probabilities for $\sigma_i(t)$ given the values of its
neighbors $\sigma_{\pm1} (t-1)$ are: $P(1|0,0)=0$, $P(1|1,0)=P(1|0,1)=p$ and 
$P(1|1,1)=q$. This model has a line of continuous phase transitions
separating the active and absorbing phases in the $p-q$ plane. For $q=p(2-p)$
the DK model is equivalent to bond directed percolation (DP), with $%
p_c=0.644700$. The critical behavior along the transition line falls in the
DP universality class, with the exception of the point $p=1/2$, $q=1$,
corresponding to so-called compact DP \cite{domany,essam}.

\section{Results}

We study the time-dependent order-parameter moments $\langle \rho
^{k}(t)\rangle $ in Monte Carlo simulations of the one-dimensional DK and
CP, obtaining time series $\left\langle \rho (t)\right\rangle _{\rho
_{0}=1/L}$ and $\left\langle \rho (t)\right\rangle _{\rho _{0}=1}$, for the
two initial conditions described above. These quantities are combined to
yield $F_{2}$ defined in Eq.(\ref{second_cumun}). Initially we studied
systems of $L=$ $2048$ sites in $N_{s}=50000$ independent realizations, each
extending to $t_{\max }=1000$. Analyzing our results, we find $z=1.5801(4)$
for the CP and $z=1.5804(2)$ for the DK. (For the CP data for times in the
interval $[30,1000]$ are used; for the DK model the interval is $[40,1000]$.
The exponent values are obtained from linear fits to the data on log
scales.) Similar results are obtained for $L=4096$. The evolution of $F_{2}$
for the two models is shown in Fig. 1. These results are in agreement with
previous results on DP ($\ z=1.580745(10)$ \cite{Jensen2}) obtained via a
low-density expansion, and on the CP ($z=1.58077(2)$) from exact
diagonalization of the master equation \cite{J. R. Mend}.

\begin{figure}[th]
\centerline{\psfig{file=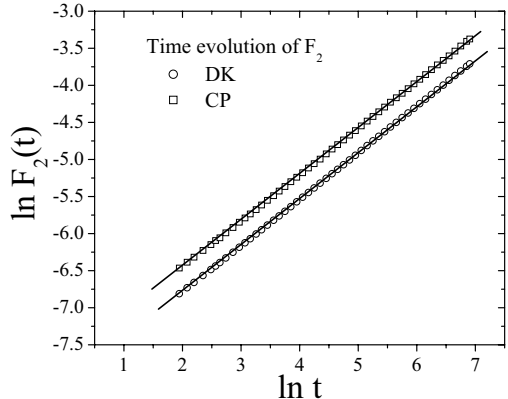,width=8cm}} \vspace*{8pt}
\caption{Time evolution of the $F_{2}$ to CP and DK.}
\end{figure}

To obtain our best estimate, we divide the time interval into subintervals
equally spaced in $\ln t$. This avoids placing an undluy large weight on
longer times, as would occur if each integer time were treated as a separate
data point.

In this case was used $N_{s}=10000$ samples for each independent seed, to a
total of 5 seeds, that gives the same $N_{s}=50000$ samples, however with
lesser errors due the statistical correlations between samples.

We study various intervals of $\ln t$ to obtain our estimates; for example,
for $\ [2.197,6.904]$ we obtain $1/z=0.62867(52)$ and $1/z=0.62802(44)$
respectively for the CP and DK. Our best estimates are found respectively
using the intervals $[3.555,6.904]$ for the CP and $[4.9397,6.904]$ for the
DK, yielding $1/z=0.63240(25)$ $(z=1.5813(6))$ and $1/z=0.63252(37)$ $(z=$\ $%
1.5810(9))$. These results are more realiable and are consistent with those
of \cite{Jensen2} and \cite{J. R. Mend}.

\begin{figure}[th]
\centerline{\psfig{file=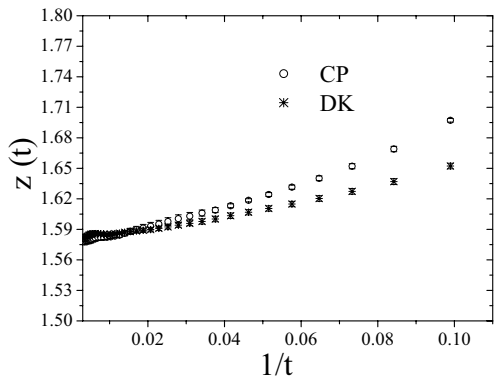,width=8cm}} \vspace*{8pt}
\caption{Effective exponent $z(t)$ versus $1/t$ to the CP and DK. Note the
convergence to value $z\approx 1.58$ in both cases.}
\end{figure}

Numerically $z_{t}$ can be estimated from a least-squares linear fit; it is
plotted versus $t_{a}^{-1}$, $t_{a}$ being the geometric mean of the $t-$%
values over the subintervals. The exponents $z_{t}$ and $\nu _{//}$ are
similarly obtained, and an extrapolation $t_{a}^{-1}\rightarrow 0$ is
performed. In Fig 2 we show this extrapolation for the CP and DK, showing
convergence towards $z\approx 1.58$, that reflects the universality between
the two models.

The ratio 
\begin{equation}
m(t)=\frac{\left\langle \rho ^{2}(t)\right\rangle _{\rho _{0}=1}}{%
\left\langle \rho (t)\right\rangle _{\rho _{0}=1}^{2}}
\end{equation}%
converges to the expected value for models in the DP universality class, $%
m_{\infty }\sim 1.1174$ to DP in $(1+1)$ dimensions \cite{rdjaff}. At short
times, $m-1=\mbox{var}(\rho )/\left\langle \rho \right\rangle ^{2}$
increases as a power law. The associated exponent is found by noting that

\begin{equation}
\chi \equiv L^{d} \mbox{var}(\rho )\sim t^{\phi }g[(p-p_{c})^{\nu _{||}}t]%
\overset{t\rightarrow \infty }{\rightarrow }(p-p_{c})^{\gamma }
\end{equation}%
so that $g(x)\sim x^{-\gamma /\nu_{||}}$ for large $x$. Observing that $\phi
=\gamma /\nu _{||}=(d\nu_{\perp }-2\beta )/\nu _{||}$, and that $%
\left\langle \rho \right\rangle \sim t^{-\beta /\nu _{_{||}}}$, we expect $%
m-1\sim t^{d/z}$. For DP in $(1+1)$ dimensions, $1/z=0.6326$. Our
simulations of the CP ($L=5000$, $N_{s}=10000)$, give $z^{-1}=0.634(3)$.

\begin{figure}[th]
\centerline{\psfig{file=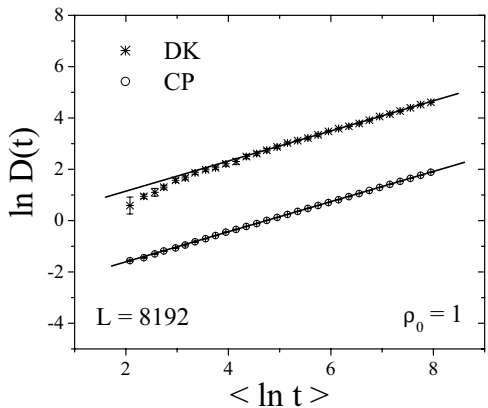,width=8cm}} \vspace*{8pt}
\caption{Plot of time evolution of $D(t)$ in the CP and DK.}
\end{figure}


In the determination of $\nu _{||}$ via Eq. (\ref{powerlawderivative}), we
used $t_{\max }=2980$. In Fig. 3 we show the derivative $D(t)$, and in Fig.
4 the effective exponent $\nu _{||}(t)$ is plotted versus $1/t$ for the CP
and DK, for the case $\rho _{0}=1$. We find $\nu _{||}(t)\approx 1.70$, as $%
t\rightarrow \infty $. We refined this estimate using lattices of $L=2048$, $%
4096$ and $8192$ for the DK model, to gauge finite-size effects. The
estimates for $\nu _{||}$ (using a total of $N_{s}=10000$ realizations, and $%
\Delta h=2\delta =$ $0.002$) were $1.623(2)$, $1.625(3)$ and $1.648(5)$
respectively. 
\begin{figure}[th]
\centerline{\psfig{file=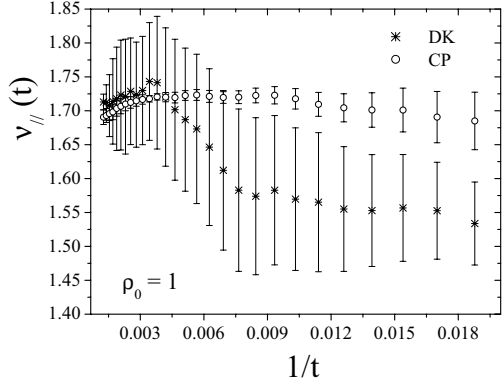,width=8cm}} \vspace*{8pt}
\caption{Effective exponent $\protect\nu _{||}(t)$ versus $1/t$ to the CP
and DK with $\protect\rho _{0}=1$.}
\end{figure}
These differ from the accepted value $\nu _{||}=1.733847(6)$ \cite%
{Dickman1,Jensen2}. Reducing $\Delta h$ by an order of magnitude, to 0.0002,
we obtain $\nu _{||}=1.731(9)$, consistent with the expected result. In the
CP simulations we used the reweighting method proposed in \cite{Dickman2}.
The sample of realizations generated at $\lambda _{c}$ is reweighted for
nearby values, $\lambda _{c}\pm n\delta $. Using $\delta =0.001$, we
determine $\nu _{||}(t)$ (figure 5) to study the influence of $\Delta h$.
The last ten points of each curve are extrapolated to $1/t\rightarrow 0$ to
estimate $\nu _{||}$; as $\delta $ is reduced, our estimate approaches the
expected value $\nu _{||}\simeq 1.73$. (For $\delta <0.001$ the curves
become indistinguishable.) The value found via extrapolation is $\nu
_{||}=1.734(8)$, in agreement with the accepted value.

\begin{figure}[th]
\centerline{\psfig{file=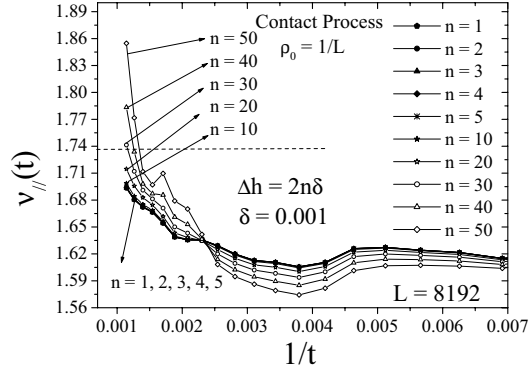,width=8cm}} \vspace*{8pt}
\caption{Effects of $\Delta h$ size in the curves of effective exponent in
the CP modelo to $\protect\rho _{0}=1/L$. Non-equilibrium reweighting and
the continuous time MC method were used.}
\end{figure}

A interesting way of measure the direct influence of $\Delta h$ in $\nu
_{||} $ can be seen in a plot of $\nu _{||}$ as function of $\Delta h$. We
note the expected behavior of $\nu _{||}\rightarrow 1.73...$. (figure 6).

\begin{figure}[th]
\centerline{\psfig{file=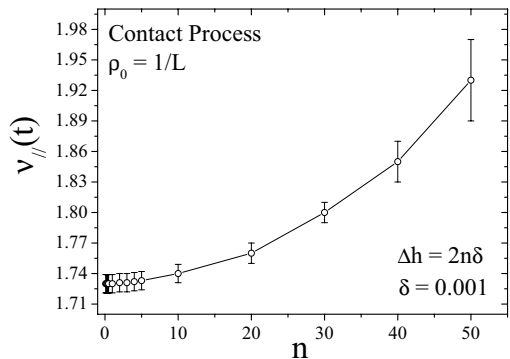,width=8cm}} \vspace*{8pt}
\caption{Plot of $\protect\nu _{||}$ as a function of $n$, where $\Delta h=2n%
\protect\delta $. }
\end{figure}

\section{Summary and Conclusions}

We apply several simulation methods based on analysis of short time
simulations to determine the critical exponents $z$ and $\nu _{||}$ in
models exhibiting a continuous phase transition to an absorbing state. The
ratios $F_{2}(t)=\left\langle \rho \right\rangle _{\rho
_{0}=1/L^{d}}/\left\langle \rho \right\rangle _{\rho _{0}=1}^{2}$ and $%
m(t)=\left\langle \rho (t)^{2}\right\rangle _{\rho _{0}=1}/\left\langle \rho
(t)\right\rangle _{\rho _{0}=1}^{2}$ have shown to be useful in alternative
methods to determine critical exponents of those models. Our estimates for
the dynamic critical exponent $z$ ($1.5813(6)$ for the CP and $1.5810(9)$
for the DK cellular automaton) are in good agreement with accepted results
in the literature obtained by other techniques such as low-density expansion
and exact diagonalization of the master equation.

In order to improve the efficiency in determining the critical exponent $%
\nu_{||}$ we employ a list of occupied sites and sample reweighting. We also
check the influence of $h$, the increment used in evaluating the numerical
derivative in Eq. (\ref{numeric derivative}). The estimates for $\nu_{||}$
appear to be very sensitive to increment size. Using $\Delta h=2\delta =$ $%
0.0002$ we obtain our best estimates which are $\nu _{||}=1.731(9)$ for the
DK cellular automata and $\nu _{||}=1.734(8)$\ for the CP process. These
results are in agreement with, though considerably less precise than, the
best estimate in the literature, $\nu _{||}=1.733847(6)$ \cite{Jensen2}. We
believe that the methods investigated here will be useful in the analysis of
other models with absorbing states, in particular, in establishing the
universality class using short time simulations, which are typically less
computationally demanding than studies of the stationary process.

\section*{Acknowledgments}

R. da Silva thanks Ricardo D. da Silva (in memorian), for all words of
incentive in its career and life and to CNPq for parcial financial support.

\end{document}